\documentclass[twocolumn,showpacs,amsmath,amssymb,prl,superscriptaddress]{revtex4}

\usepackage{graphicx,here}
\usepackage{dcolumn}
\usepackage{bm}
\usepackage{enumerate}

\begin{document}

\preprint{APS/123-QED}

\title{Electric Polarization Induced by a Proper Helical Magnetic Ordering\\ %
in a Delafossite Multiferroic CuFe$_{1-x}$Al$_x$O$_2$}



\author{T. Nakajima}
\email{E-mail address: nakajima@nsmsmac4.ph.kagu.tus.ac.jp}
\author{S. Mitsuda}
\author{S. Kanetsuki}
\author{K. Tanaka}
\author{K. Fujii}
\affiliation{Department of Physics, Faculty of Science, Tokyo University of Science, Tokyo 162-8601, Japan}%
\author{N. Terada}
\affiliation{ICYS, National Institute for Materials Science, Ibaraki 305-0044, Japan}%
\author{M. Soda}
\author{M. Matsuura}
\author{K. Hirota}
\affiliation{Institute for Solid State Physics, University of Tokyo, Kashiwa 277-8581, Japan}%

\begin{abstract}
Multiferroic CuFe$_{1-x}$Al$_x$O$_2$ $(x=0.02)$ exhibits a ferroelectric ordering accompanied %
by a proper helical magnetic ordering below $T=7$K under zero magnetic field. %
By polarized neutron diffraction and pyroelectric measurements, we have 
revealed a one-to-one correspondence between the spin helicity and the direction of the spontaneous electric polarization. %
This result indicates that the spin helicity of the proper helical magnetic ordering %
is essential for the ferroelectricity in CuFe$_{1-x}$Al$_x$O$_2$. %
%
%
%
The induction of the electric polarization by the proper helical magnetic ordering is, however, cannot be explained %
by the Katsura-Nagaosa-Balatsky model, %
%
which successfully explains the ferroelectricity in the recently explored ferroelectric helimagnets, such as TbMnO$_3$. %
%
%
We thus conclude that CuFe$_{1-x}$Al$_x$O$_2$ is a new class of magnetic ferroelectrics. %
\end{abstract}

\date{July 13, 2007, {\bf preprint}}

\maketitle

Novel types of couplings between dielectric property and magnetism, which produce colossal %
magnetoelectric (ME) effects, have been extensively investigated %
since a gigantic ME effect was discovered in $R$MnO$_3$ ($R$ is a rare earth material) \cite{Kimura_nature}. %
%
Among several types of couplings between spins and electric polarizations, %
a ferroelectricity induced by noncollinear spin arrangements has been most widely investigated %
experimentally and theoretically \cite{Katsura_PRL_2005,Mostovoy_PRL_2006,Kenzelmann_PRL_Spiral,Arima_PRL_Spiral,PRL_Ni3V2O8,PRL_MnWO4,Yamasaki_PRL_conical}. %
Katsura, Nagaosa and Baratsky (KNB) %
proposed that the local electric dipole moment $\mbox{\boldmath $p$}$, which arises between neighboring two spins %
$\mbox{\boldmath $S$}_i$ and $\mbox{\boldmath $S$}_{i+1}$, can be described in the form of %
$\mbox{\boldmath $p$}\propto\mbox{\boldmath $e$}_{i,i+1}\times (\mbox{\boldmath $S$}_i \times \mbox{\boldmath $S$}_{i+1})$, %
where $\mbox{\boldmath $e$}_{i,i+1}$ is a unit vector connecting two spins \cite{Katsura_PRL_2005}. %
This formula successfully explains the ferroelectric property in cycloidal or conical magnetic orderings of some %
transition metal oxides, such as $R$MnO$_3$ ($R$=Tb, Tb$_{1-x}$Dy$_x$), Ni$_3$V$_2$O$_8$, MnWO$_4$ %
and CoCr$_2$O$_4$ \cite{Kenzelmann_PRL_Spiral,Arima_PRL_Spiral,PRL_Ni3V2O8,PRL_MnWO4,Yamasaki_PRL_conical}. %
Moreover, a recent polarized neutron diffraction study on TbMnO$_3$ demonstrated that %
the spin helicity, clockwise or counterclockwise, correlates %
with the direction of the electric polarization, as predicted in the formula \cite{PRL_Helicity_TbMnO3}. %
It is, however, recently reported that ferroelectricity in a helical magnetic ordering of a delafossite multiferroic CuFe$_{1-x}$Al$_x$O$_2$ %
cannot be explained by the above formula \cite{SpinNoncollinearlity}. %
Therefore, CuFe$_{1-x}$Al$_x$O$_2$ provides an opportunity to explore an another type of spin-polarization coupling. %
\begin{figure}[b]
\begin{center}
\includegraphics[keepaspectratio,width=7.8cm,clip]{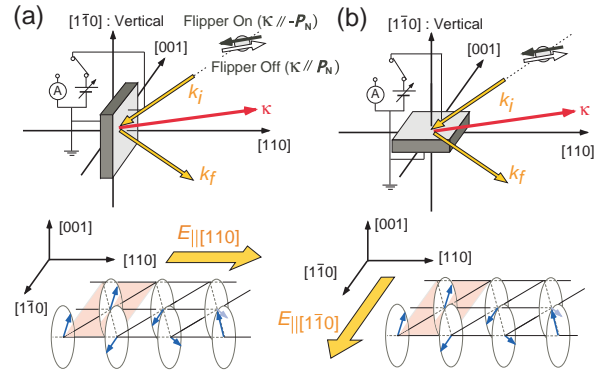}
	\caption{(Color online) Schematic illustrations of the experimental configurations and the relationship %
between the direction of the poling electric field $E$ and the proper helical magnetic structure in the FEIC phase for %
(a) the $E_{\parallel[110]}$ sample and (b) the $E_{\parallel[1\bar{1}0]}$ sample. %
}
\label{config}
\end{center}
\end{figure}

CuFeO$_2$, which is one of model materials of a triangular lattice antiferromagnet, %
has been extensively investigated as a geometrically frustrated spin system for last fifteen years %
\cite{Mitsuda_1991,Mitsuda_2000,Petrenko_2000}. %
The ground state of CuFeO$_2$ is a collinear commensurate 4-sublattice ($\uparrow\uparrow\downarrow\downarrow$) state %
with the magnetic moments along the $c$ axis, which is normal to the triangular lattice layers, %
in spite of the Heisenberg spin character expected from the electronic configuration of Fe$^{3+}$ ($S=\frac{5}{2}$, $L=0$). %
When a magnetic field is applied along the $c$ axis at low temperature, %
CuFeO$_2$ exhibits a multi-step magnetization process consisting of several magnetization plateaus and slopes, %
which is accompanied by stepwise changes of lattice constants \cite{Terada_LatticeStaircase,Terada_LatticeStaircase_FP}. %
Since Kimura and co-workers discovered a spontaneous electric polarization %
in the first field-induced phase of CuFeO$_2$, %
which emerges along the direction perpendicular to the $c$ axis \cite{Kimura_CuFeO2}, %
CuFeO$_2$ has also been investigated as a candidate of novel multiferroic materials. %
Recent studies on the slightly diluted system CuFe$_{1-x}$Al$_x$O$_2$ %
showed that only a few percent dilution of Fe$^{3+}$ sites with nonmagnetic Al$^{3+}$ ions %
considerably reduces the transition field from the 4-sublattice phase to the field-induced ferroelectric phase. %
Moreover, the ferroelectric phase shows up even under zero field %
in the concentration region of $0.014<x<0.030$ \cite{Kanetsuki_JPCM,Seki_PRB_2007,Terada_x_T}. %
Quite recently, the magnetic structure in the ferroelectric phase was  elucidated to be %
an antiferromagnetically-stacked proper helical structure with an incommensurate propagation wave vector $(q,q,0)$ %
where $q\sim0.21$ \cite{SpinNoncollinearlity}. %
In this letter, we refer to this ferroelectric phase as ferroelectric incommensurate (FEIC) phase. %
This magnetic structure, however, cannot lead to a finite uniform electric polarization through the formula %
$\mbox{\boldmath $p$}\propto\mbox{\boldmath $e$}_{i,i+1}\times (\mbox{\boldmath $S$}_i \times \mbox{\boldmath $S$}_{i+1})$, %
because the direction of $\mbox{\boldmath $e$}_{i,i+1}$ is parallel to the direction of %
$\mbox{\boldmath $S$}_i \times \mbox{\boldmath $S$}_{i+1}$ in average. %
Nevertheless, %
the spin helicity, a right-handed (RH) or left-handed (LH) proper helical arrangement of spins, %
is expected to correlate with the direction of the electric polarization, %
because space inversion operation flips the spin helicity, as well as the direction of electric polarization. 
%
In present work, %
we thus performed polarized neutron diffraction %
and pyroelectric measurements using CuFe$_{1-x}$Al$_x$O$_2$ samples %
with $x=0.02$, 
which exhibits the ferroelectric ordering below $T=7$K, %
in order to elucidate the relationship between the spin helicity and the electric polarization. %

\begin{figure}[t]
\begin{center}
\includegraphics[keepaspectratio,width=7.8cm,clip]{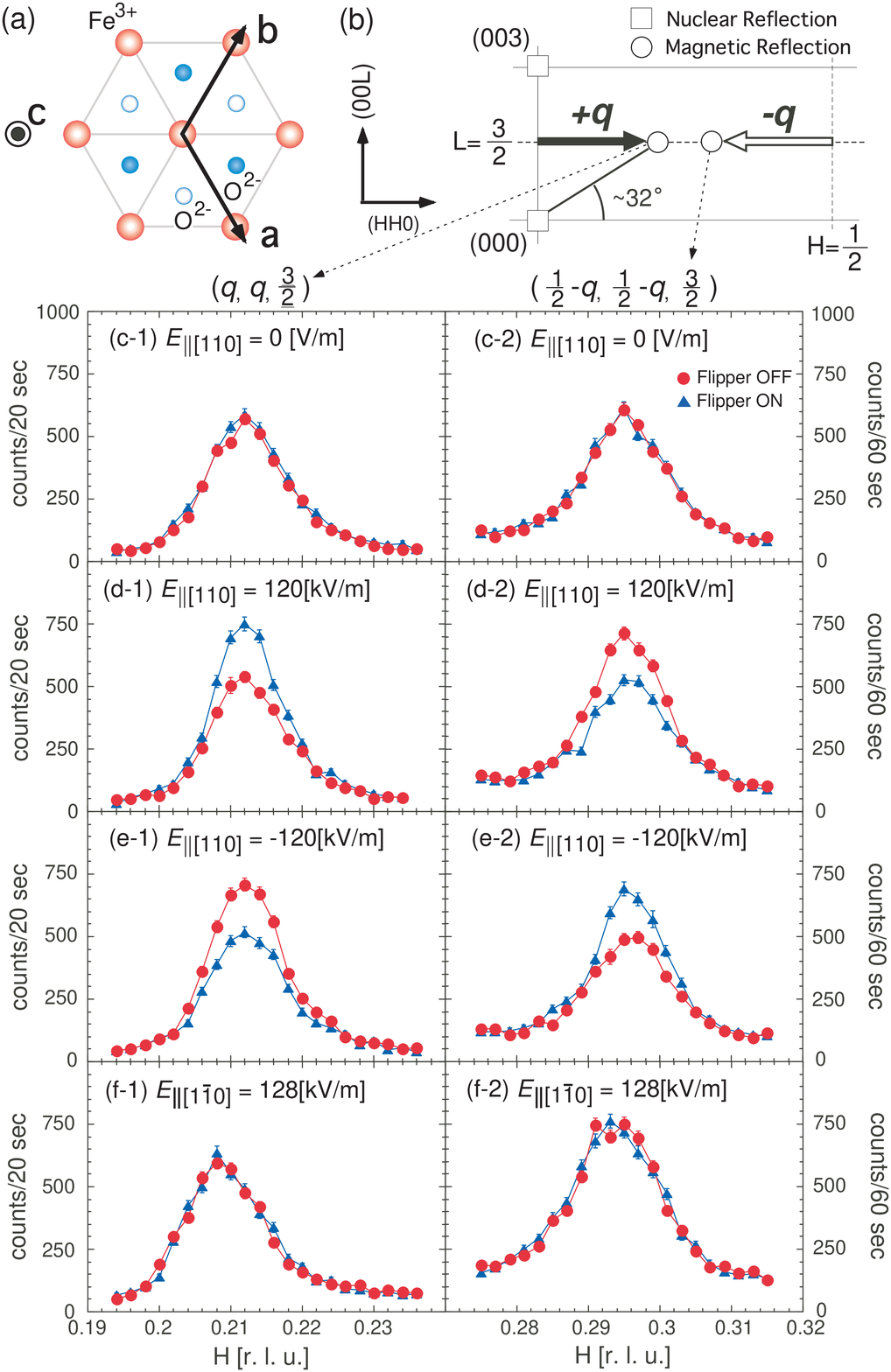}
	\caption{(Color online) (a) The the hexagonal basis represented on the Fe$^{3+}$ triangular lattice layer. %
	Open and 	filled blue circles denote O$^{2-}$ ions located above and below the Fe$^{3+}$ layer, respectively. %
	(b) The location of the magnetic reflections surveyed in present measurement in $(HHL)$ zone. %
	(c-1)-(f-2) The diffraction profiles of $(H,H,\frac{3}{2})$ reciprocal lattice scans for the $(q,q,\frac{3}{2})$ and $(\frac{1}{2}-q,\frac{1}{2}-q,\frac{3}{2})$ magnetic Bragg %
	reflections at $T=2$K in the FEIC phase. %
}
\label{profiles}
\end{center}
\end{figure}

A single crystal of CuFe$_{1-x}$Al$_x$O$_2$ with $x=0.02$ of nominal composition was %
prepared by the floating zone technique \cite{Zhao_FZ}, and cut into two pieces with disk shapes; %
one of them has the widest surface normal to the [110] axis ($E_{\parallel[110]}$ sample), %
the other has that normal to the $[1\bar{1}0]$ axis ($E_{\parallel[1\bar{1}0]}$ sample). %
The experimental configurations for these samples are illustrated in Figs. \ref{config}(a) and (b). %
Silver paste was pasted on the widest surface of each sample to make the electrodes. %
The polarized neutron diffraction measurements were carried out at the triple-axis neutron %
spectrometer PONTA installed by University of Tokyo at JRR-3 in Japan Atomic Energy Agency. %
The incident polarized neutron with the energy of 34.05 meV was obtained %
by a Heusller (111) monochromator. %
The flipping ratio of the polarized neutron beam was $19.0$, %
and the polarization vector of the incident neutron, $\mbox{\boldmath $p$}_{\rm N}$, was set to be parallel (or antiparallel) %
to the scattering vector, $\mbox{\boldmath $\kappa$}$, by a guide-field of a helmholtz coil and a spin flipper. %
The collimation was $40'$-$40'$-$40'$-$40'$, and a pyrolytic graphite analyzer was employed. %
The sample was mounted in a pumped $^4$He cryostat with the %
($hhl$) scattering plane. 
Note that, in the present experiment, we employed a conventional hexagonal basis as was in the previous works, %
while CuFe$_{1-x}$Al$_x$O$_2$ originally has a trigonal (rhombohedral) crystal structure. %
The definition of the hexagonal basis is shown in Fig. \ref{profiles}(a). %
For the measurements of the spontaneous electric polarization $\mbox{\boldmath $P$}$, pyroelectric current was measured %
under zero electric field with increasing temperature, using an electrometer (Keithley 6517A). %
Before each neutron diffraction (or pyroelectric) measurement, we performed a proper cooling %
with applied electric field from 20K to 2K. %

\begin{figure*}[t]
\begin{center}
\includegraphics[keepaspectratio,width=14.5cm,clip]{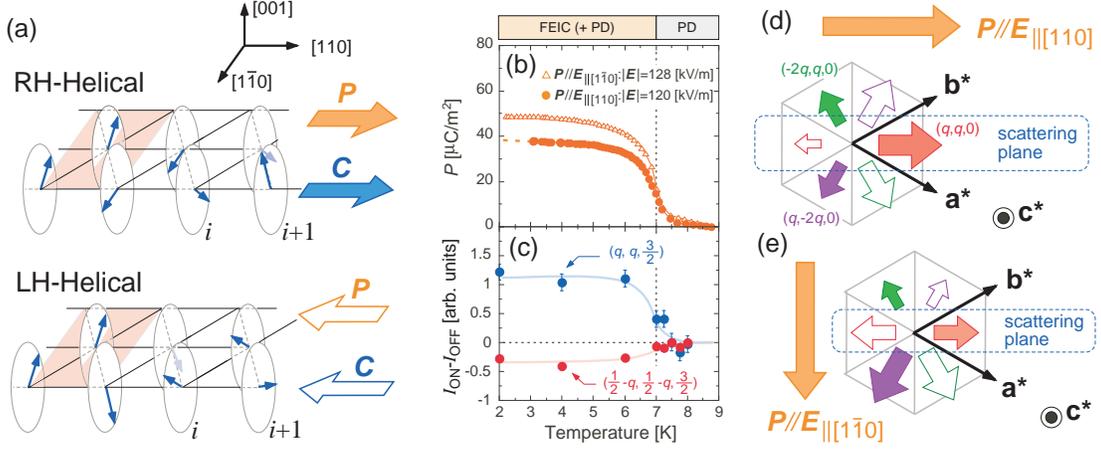}
	\caption{(Color online) (a) The relationship between the spin helicity and the direction of the electric polarization. %
	(b) The temperature variations of the electric polarization along the $[110]$ and $[1\bar{1}0]$ axes %
	measured after cooling with the poling electric field parallel to the $[110]$ and $[1\bar{1}0]$ directions, respectively. %
	(c) The temperature dependence of $I_{\rm ON}-I_{\rm OFF}$ measured on heating after a cooling with a poling electric field %
	(120 kV/m) parallel to $[110]$ axis. %
	The schematic drawings of the distributions of the RH- (filled arrows) and LH- (open arrows) helical orderings %
	among magnetic domains with three equivalent propagation wave vectors $(q,q,0), (-2q,q,0)$ and $(q,-2q,0)$, %
	when the macroscopic electric polarization emerges along (d) the $[110]$ axis %
	and (e) the $[1\bar{1}0]$ axis. The fractions of the RH- (or LH-) helical orderings are represented by the sizes of arrows. %
	}
\label{Helical}
\end{center}
\end{figure*}

Before discussing the results of the present measurements, we briefly review the scattering cross section for polarized neutrons. %
Let us assume that scattering system consists of RH- and LH-proper helical magnetic orderings with a propagation %
wave vector $\mbox{\boldmath $q$}$. According to the Blume's notation \cite{BlumeNotation}, the scattering cross section for %
a pair of magnetic satellite reflections located at $\mbox{\boldmath $\tau$}\pm\mbox{\boldmath $q$}$, %
where $\mbox{\boldmath $\tau$}$ is a reciprocal lattice vector, is described as follows: %
\begin{eqnarray}
\bigg(\frac{d\sigma}{d\Omega}\bigg)_{\mbox{\boldmath $\tau$}\pm\mbox{\boldmath $q$}} %
\propto & S(\mbox{\boldmath $\kappa$})\{(1+(\mbox{\boldmath $\hat{C}$}\cdot\mbox{\boldmath $\hat{\kappa}$})^2)(V_{\rm RH}+V_{\rm LH})\nonumber\\ %
 & \mp 2(\mbox{\boldmath $p$}_{\rm N}\cdot\mbox{\boldmath $\hat{\kappa}$})(\mbox{\boldmath $\hat{C}$}\cdot\mbox{\boldmath $\hat{\kappa}$})(V_{\rm RH}-V_{\rm LH})\},%
\end{eqnarray}
where $S(\mbox{\boldmath $\kappa$})$ is the factor depending on the magnetic structure factor, %
$V_{\rm RH}$ and $V_{\rm LH}$ are the volumes of the RH- and LH-helical orderings, respectively, %
$\mbox{\boldmath $\hat{\kappa}$}$ is a unit vector of $\mbox{\boldmath $\kappa$}$, $|\mbox{\boldmath $p$}_{\rm N}|=1$, and %
$\mbox{\boldmath $\hat{C}$}$ is a unit vector corresponding to the spin helciity, %
(refered as `vector spin chirality' in Ref.\cite{PRL_Helicity_TbMnO3}) which is defined so that $\mbox{\boldmath $S$}_i,\mbox{\boldmath $S$}_{i+1}$ and $\mbox{\boldmath $C$}$ in this order form a right-handed coordinate system (see Fig. \ref{Helical}(a)). %
Note that the above expression of the scattering cross section includes both the `spin-flip' and `non spin-flip' scatterings, %
and thus polarization analysis for scattered neutrons is not necessary. %
In the  present experiment, %
$\mbox{\boldmath $C$}$ is parallel (antiparallel) to the $[110]$ direction for the RH- (LH-) proper helical ordering, %
and the cross sections of two magnetic Bragg reflections at $(q,q,\frac{3}{2})$ and $(\frac{1}{2}-q,\frac{1}{2}-q,\frac{3}{2})$, %
which are mainly surveyed in the present measurements, %
correspond to $(d\sigma/d\Omega)_{\mbox{\boldmath $\tau$}+\mbox{\boldmath $q$}}$ and $(d\sigma/d\Omega)_{\mbox{\boldmath $\tau$}-\mbox{\boldmath $q$}}$, respectively (see Fig. \ref{profiles}(b)). %
In this case, the imbalance between $V_{\rm RH}$ and $V_{\rm LH}$ is expressed as following: %
\begin{eqnarray}
\frac{V_{\rm RH}-V_{\rm LH}}{V_{\rm RH}+V_{\rm LH}}%
&=& %
A(\mbox{\boldmath $\kappa$})%
\bigg(\frac{I_{\rm ON}-I_{\rm OFF}}{I_{\rm ON}+I_{\rm OFF}}\bigg),
\label{A_kappa}
\end{eqnarray}
where $I_{\rm ON}$ and $I_{\rm OFF}$ are the intensities of a magnetic Bragg reflection %
measured when the spin flipper is on $(\mbox{\boldmath $p$}_{\rm N}\parallel -\mbox{\boldmath $\kappa$})$ %
and off $(\mbox{\boldmath $p$}_{\rm N}\parallel \mbox{\boldmath $\kappa$})$, respectively. %
The values of the proportional constant $A(\mbox{\boldmath $\kappa$})$ %
for the $(q,q,\frac{3}{2})$ and $(\frac{1}{2}-q,\frac{1}{2}-q,\frac{3}{2})$ %
magnetic reflections are approximately $1$ and $-1$, respectively. %

In Fig. \ref{profiles}, we now show typical diffraction profiles of magnetic reflections in the FEIC phase ($T=2$K). %
After cooling the $E_{\parallel[110]}$ sample under zero electric field, %
as shown in Figs. \ref{profiles} (c-1) and (c-2), there was no difference between %
$I_{\rm ON}$ and $I_{\rm OFF}$ for both of the $(q,q,\frac{3}{2})$ and the $(\frac{1}{2}-q,\frac{1}{2}-q,\frac{3}{2})$ reflections. %
This result indicates that %
the fractions of the RH- and LH-helical magnetic orderings were equal to each other. %
After cooling the $E_{\parallel[110]}$ sample under a poling electric field (120 kV/m) parallel to the $[110]$ direction, %
$I_{\rm ON}$ was greater than $I_{\rm OFF}$ for the $(q,q,\frac{3}{2})$ reflection, %
and this relationship between $I_{\rm ON}$ and $I_{\rm OFF}$ was reversed for the %
$(\frac{1}{2}-q,\frac{1}{2}-q,\frac{3}{2})$ reflection, as shown in Figs. \ref{profiles}(d-1) and (d-2). %
%
%
By a reversal of the direction of the poling electric field applied on cooling, %
this relationship between $I_{\rm ON}$ and $I_{\rm OFF}$ %
for each magnetic satellite was reversed, as shown in Figs. \ref{profiles}(e-1) and (e-2). %
No imbalance between $I_{\rm ON}$ and $I_{\rm OFF}$, however, was %
observed for the  $E_{\parallel[1\bar{1}0]}$ sample %
after cooling under the poling electric field (128 kV/m) parallel to the $[1\bar{1}0]$ direction, %
as shown in Figs. \ref{profiles}(f-1) and (f-2). %
These results show that the poling electric field along the $[110]$ axis %
induces an imbalance between the fractions of the RH- and LH-helical orderings, %
but the poling electric field along $[1\bar{1}0]$ axis does not. %
Taking account of the fact that %
the poling electric field along the $[110]$ axis also induces %
the macroscopic electric polarization along the $[110]$ axis (see Fig. \ref{Helical}(b)), %
we conclude that a proper helical magnetic ordering %
generates an electric polarization along the helical axis, %
and moreover, there is the one-to-one correspondence between 
the spin helicity and the direction of electric polarization, as illustrated in Fig. \ref{Helical}(a). %

\begin{figure}[t]
\begin{center}
\includegraphics[keepaspectratio,width=7.8cm,clip]{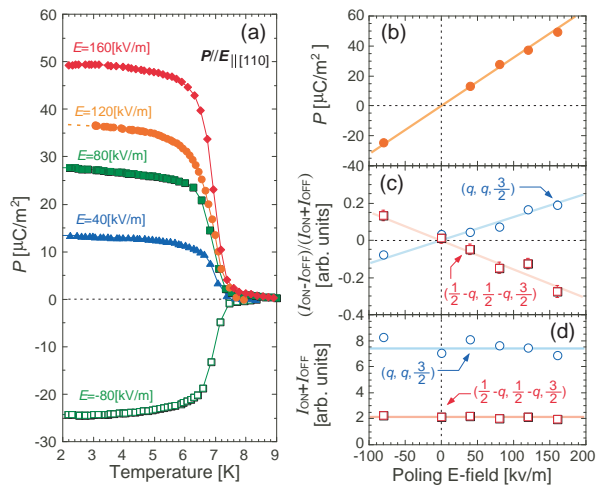}
	\caption{(Color online) (a) Temperature variations of the electric polarization along the $[110]$ axis measured after cooling under various poling electric fields along the $[110]$ axis. 
	The poling electric field dependence of (b) the spontaneous electric polarization, %
	(c) $(I_{\rm ON}-I_{\rm OFF})/(I_{\rm ON}+I_{\rm OFF})$ and (d) $(I_{\rm ON}+I_{\rm OFF})$ at $T=2$K. %
	The colored solid lines are guides to eyes.}%
\label{pyro}
\end{center}
\end{figure}

Although the poling electric field along the $[1\bar{1}0]$ axis also induces %
the macroscopic electric polarization along the $[1\bar{1}0]$ axis, as shown in Fig. \ref{Helical}(b), %
this can be ascribed to the existence of three magnetic domains reflecting %
the trigonal three-fold symmetry of the crystal structure (see Fig. \ref{profiles}(a)). %
When the electric polarization emerges along the $[1\bar{1}0]$ axis, %
the imbalance between the fractions of the RH- and LH-helical ordering must be induced %
in the domains out of the scattering plane, as illustrated in Fig. \ref{Helical}(e). %
Figs. \ref{pyro}(a)-(c) show %
the poling electric field dependence of the spontaneous electric polarization %
and the imbalance between $I_{\rm ON}$ and $I_{\rm OFF}$ for the $E_{\parallel(110)}$-sample. %
We found that both the %
magnitude of the electric polarization and the imbalance between $I_{\rm ON}$ and $I_{\rm OFF}$ at $T=2$K %
are proportional to the poling electric field applied on cooling. %
In addition, the temperature variation of $I_{\rm ON}-I_{\rm OFF}$, which was measured %
in a warming rum under zero electric field after a cooling with the poling electric field parallel to the $[110]$ axis, %
is similar to that of the spontaneous electric polarization, as shown in Figs. \ref{Helical}(c). %
These results apparently show that the the macroscopic electric polarization %
arises from the `imbalance' between the fractions of the RH- and LH-helical magnetic orderings. %
Note that the fractions of the RH- and LH-helical magnetic orderings, however, %
cannot be determined accurately in the present experiments, %
because the thermally induced partially disordered (PD) state, which has a collinear incommensurate magnetic structure %
with almost the same wave number as that of the FEIC magnetic ordering, %
is supposed to remain even in the FEIC phase owing to the pinning effect by nonmagnetic impurity ions \cite{SpinNoncollinearlity,Terada_x_T}. %
While the imbalance between $I_{\rm ON}$ and $I_{\rm OFF}$ apparently shows the poling electric field dependence %
as mentioned above, %
the sum of $I_{\rm ON}$ and $I_{\rm OFF}$ does not, %
as shown in Fig. \ref{pyro}(d). %
This means that the application of a poling electric field within $|E|<\sim160$ kV/m %
does not affect the fractions of the three magnetic domains, %
whose propagation wave vectors are different to each other. %

In summary, %
we performed polarized neutron diffraction and pyroelectric measurements %
on the delafossite multiferroic CuFe$_{1-x}$Al$_x$O$_2$ with $x=0.02$, %
and demonstrated that %
the proper helical magnetic ordering of CuFe$_{1-x}$Al$_x$O$_2$ generates %
a spontaneous electric polarization parallel to the helical axis. %
This indicates that the local spin-polarization coupling in CuFe$_{1-x}$Al$_x$O$_2$ %
cannot be explained by the KNB-model %
($\mbox{\boldmath $p$}\propto\mbox{\boldmath $e$}_{i,i+1}\times (\mbox{\boldmath $S$}_i \times \mbox{\boldmath $S$}_{i+1})$). %
Nevertheless, the results of the present study revealed %
a one-to-one correspondence between the spin helicity and the direction of the electric polarization, %
indicating that the spin helicity of the proper helical magnetic ordering is essential for the ferroelectricity in CuFe$_{1-x}$Al$_x$O$_2$. %
Quite recently, Arima proposed that a proper helical magnetic order can generate %
ferroelectricity through the variation in the metal-ligand hybridization with spin-orbit coupling \cite{Arima_Symmetry}. %
The present results suggest that this mechanism is applicable to CuFe$_{1-x}$Al$_x$O$_2$. %
We thus conclude that CuFe$_{1-x}$Al$_x$O$_2$ is a new class of magnetic ferroelectrics, %
which will pave another way to design multiferroic materials. %

We are grateful to T. Arima for fruitful discussions. %
The neutron diffraction measurement at JRR-3M was supported by ISSP of the University %
of Tokyo (PACS No. 7546B (5G:PONTA)). 
This work was supported by a Grant-in-Aid for Scientific Research (C), No. 19540377, from JSPS, Japan. %


\end{document}